\documentclass{article}
\usepackage{ijcai19}
\usepackage{threeparttable}
\usepackage{xcolor}
\usepackage{subfigure}
\usepackage{amsthm}
\usepackage{epsfig,algorithmic,algorithm}
\usepackage{subfigure}
\usepackage{multirow}
\usepackage{bm}
\usepackage{amssymb}
\usepackage[english]{babel}
\usepackage{arydshln}
\usepackage{times}
\usepackage{soul}
\usepackage{url}
\usepackage[hidelinks]{hyperref}
\usepackage[utf8]{inputenc}
\usepackage[small]{caption}
\usepackage{graphicx}
\usepackage{amsmath}
\usepackage{booktabs}
\title{Deep Session Interest Network for Click-Through Rate Prediction}

\author{
Yufei Feng$^1$\and
Fuyu Lv$^1$\and
Weichen Shen$^1$\And
Menghan Wang$^{1,2}$\And
Fei Sun$^{1}$\And
Yu Zhu$^{1}$\And
Keping Yang$^{1}$\\
\affiliations
$^1$Alibaba Group, Hangzhou, China\\
$^2$Zhejiang University, China\\
\emails
fyf649435349@gmail.com,
\{fuyu.lfy, weichen.swc, xiangyu.wmh, ofey.sf, zy143829\}@alibaba-inc.com, 
shaoyao@taobao.com
}

\begin{document}

\maketitle

\begin{abstract}
Click-Through Rate (CTR) prediction plays an important role in many industrial applications, such as online advertising and recommender systems.
How to capture users' dynamic and evolving interests from their behavior sequences remains a continuous research topic in the CTR prediction.
However, most existing studies overlook the intrinsic structure of the sequences: the sequences are composed of sessions, where sessions are user behaviors separated by their occurring time. We observe that user behaviors are highly homogeneous in each session, and heterogeneous cross sessions.
Based on this observation, we propose a novel CTR model named Deep Session Interest Network (DSIN) that leverages users' multiple historical sessions in their behavior sequences. We first use self-attention mechanism with bias encoding to extract users' interests in each session. Then we apply Bi-LSTM to model how users' interests evolve and interact among sessions. Finally, we employ the local activation unit to adaptively learn the influences of various session interests on the target item. Experiments are conducted on both advertising and production recommender datasets and DSIN outperforms other state-of-the-art models on both datasets.
\end{abstract}

\section{Introduction}
Recommender systems (RS) are becoming increasingly indispensable in assisting users to find their preferred items in web-scale applications such as Amazon and Taobao. Typically, an industrial recommender system consists of two stages: candidate generation and candidate ranking \cite{Paul:YoutubeNet}. The candidate generation stage adopts some naive but time-efficient recommendation algorithms (e.g. item-based collaborative filtering \cite{sarwar2001item}) to provide a relative small set of items from the huge whole item set for ranking. In the candidate ranking stage, complex but powerful models (e.g. neural network methods) are applied to rank the candidates so as to select the top-k items for recommendation. In this paper, we mainly focus on the candidate ranking stage and treat it as a Click-Through Rate (CTR) prediction task.
It means we assume a relative small item set has been provided for ranking and we rank items according to their CTR score predictions.

\begin{figure}[tb]
\begin{center}
\includegraphics[scale=0.39]{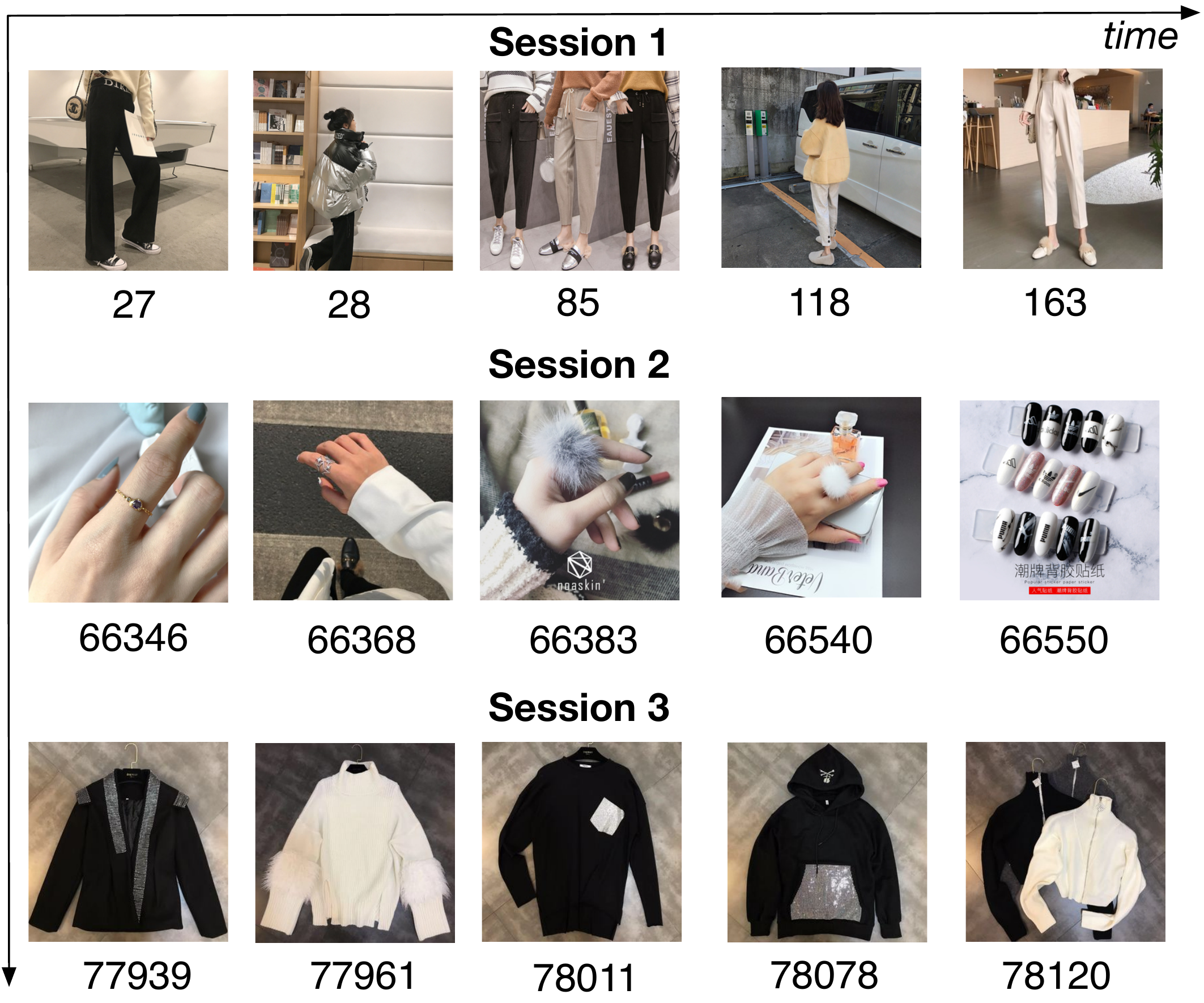}
\end{center}
   \caption{Here is one demo of sessions collected from a real-world industrial application. The number beneath the picture indicates the time gap, specified in seconds, between the time of clicking on the current item and that of clicking on the first item. Sessions are divided in the principle of whenever there exists a time gap of more than 30 minutes.}
\label{fig:Session_Demo2}
\end{figure}

Some recent effective CTR models \cite{Paul:YoutubeNet,Zhou:DIN,zhou2018deep,zhou2018atrank} show promising results by utilizing users' sequential behaviors, which reflect users' dynamic and evolving interests.
However, these models overlook the intrinsic structure of the sequences: the sequences are composed of sessions. A session is a list of interactions (user behaviors) that occur within a given time frame. We observe that user behaviors are highly homogeneous in each session and heterogeneous cross sessions.
As shown in figure \ref{fig:Session_Demo2}, a user is sampled from a real-world industrial application and we split her behavior sequence into 3 sessions. Sessions are divided in the principle of whenever there exists a time gap of more than 30 minutes \cite{Mihajlo:airbnb}.
The user mainly browses trousers in session 1, finger rings in session 2, and coats in session 3. The phenomenon illustrated in figure 1 is general. It reflects the fact that a user usually has a clear unique intention in one session while his/her interest can change sharply when he/she starts a new session.

Motivated by the above observations, we propose Deep Session Interest Network\footnote{https://github.com/shenweichen/DSIN} (DSIN) to model users’ sequential behaviors in the CTR prediction task by leveraging their multiple historical sessions.
There are three key components in DSIN. First, we naturally divide users' sequential behaviors into sessions and then use self-attention network with bias encoding to model each session. Self-attention can capture the inner interaction/correlation of session behaviors and then extract users' interests of each session. These various session interests may be correlated with each other and even follow a sequential pattern \cite{Quadrana2017:Personalizing}. So in the second part, we apply bi-directional LSTM (Bi-LSTM) to capture the interaction and evolution of users' varying historical session interests. Because various session interests have different influences on the target item, finally we design the local activation unit to aggregate them w.r.t. the target item to form the final representation of the behavior sequence.

The main contributions of this paper are summarized as follows:

\begin{itemize}
    \item We highlight that user behaviors are highly homogeneous in each session and heterogeneous cross sessions, and propose a new model named DSIN, which can effectively model the user's multiple sessions for the CTR prediction.
    \item We design a self-attention network with bias encoding to get accurate interest representation of each session. Then we employ Bi-LSTM to capture the sequential relationship among historical sessions. At last, we employ the local activation unit for aggregation considering the influences of different session interests on the target item.
    \item Two groups of comparative experiments are conducted on both advertising and production recommender datasets. The experiment results demonstrate the superiority of our proposed DSIN compared with other state-of-the-art models in the CTR prediction task.
\end{itemize}

The organization of the remaining parts of this paper is as follows. Section 2 introduces some related work. Section 3 gives the detailed description of our DSIN model. Section 4 presents our experiment results and analyses on both advertising and recommender datasets.

\section{Related Work}
In this section, we mainly introduce existing studies of the CTR prediction and session-based recommendation.

\subsection{Click-Through Rate Prediction}
Recent CTR models mainly pay attention to the interaction between features. Wide\&Deep \cite{Cheng:WDL} combines the linear representation of features. DeepFM \cite{guo2017deepfm} learns the second-order crossover of features and DCN \cite{DCN} applies a multi-layer residual structure to learn higher-order representation of features. AFM \cite{xiao2017attentional} argues that not all feature interactions are equally predictive and uses attention mechanism to automatically learn weights of cross-features. To sum up, the higher-order representation and interaction of features significantly improve the expressive ability of features and the generalization ability of models.

Users' sequential behaviors imply users' dynamic and evolving interests and have been widely proven effective in the CTR prediction task. 
YoutubeNet \cite{Paul:YoutubeNet} transforms embeddings of users' watching lists into a vector of fixed length by average pooling. Deep Interest Network (DIN) \cite{Zhou:DIN} uses attention mechanism to learn the representation of users' historical behaviors w.r.t. the target item. ATRANK \cite{zhou2018atrank} proposes an attention-based framework modeling the influence between users' heterogeneous behaviors. Deep Interest Evolution Network (DIEN) \cite{zhou2018deep} uses auxiliary loss to adjust the expression of current behavior to the next behavior and then models the specific interest evolving process for different target items with AUGRU. Modeling users' sequential behaviors enriches the representation of the user and improves the prediction accuracy significantly.

\subsection{Session-based Recommendation}
The concept of session is commonly mentioned in sequential recommendation but rare in the CTR prediction task. Session-based recommendation benefits from the dynamic evolving of users' interests in sessions. General Factorization Framework (GFF) \cite{Balazs:GFF} uses sum pooling of items to represent a session. Each item has two kinds of representations, one represents itself and the other represents the context of the session. Recently, RNN-based approaches \cite{hidasi:session,hidasi2016parallel,li2018learning} are applied into session-based recommendations to capture the order relationship within a session. Based on that, \cite{li2017neural} proposes a novel attentive neural networks framework (NARM) to model the user’s sequential behavior and capture the user’s main purpose in the current session. Hierarchical RNN \cite{Quadrana2017:Personalizing} is proposed to relays end evolves latent hidden states of the RNNs across users’ historical sessions. Besides RNNs, \cite{liu2018stamp,kang2018self} apply only self-attention based models to effectively capture long-term and short-term interests of a session. \cite{tang2018personalized} uses convolutional neural network and \cite{chen2018sequential} adopts user memory network to enhances the expressiveness of the sequential model. 


\section{Deep Session Interest Network}
In this section, we introduce the Deep Session Interest Network (DSIN) in detail. We first introduce the basic deep CTR model named BaseModel, then the technical designs of DSIN that model the extraction and interaction of users' session interests. 

\subsection{BaseModel}
In this section, we mainly introduce feature representation, embedding, MLP and loss function in BaseModel.

\subsubsection{Feature Representation}
Informative features count a great deal in the CTR prediction task. Overall, we use three groups of features in BaseModel: {\itshape User Profile}, {\itshape Item Profile} and {\itshape User Behavior}. Each group consists of some sparse features: {\itshape User Profile} contains {\itshape gender}, {\itshape city}, etc.; {\itshape Item Profile} contains {\itshape seller id}, {\itshape brand id}, etc.; {\itshape User Behavior} contains the {\itshape item ids} of items that the user recently clicked on. Note that the side information of the item can be concatenated to represent itself.

\subsubsection{Embedding}
Embedding is a common technique which transforms large-scale sparse features into low-dimensional dense vectors. Mathematically, sparse features can be represented by $\textbf{E} \in \mathbb{R}^{M \times d_{model}}$ respectively, where $M$ is the size of sparse features and $d_{model}$ is the embedding size. With embedding, {\itshape User Profile} can be represented by $\textbf{X}^U \in \mathbb{R}^{N_u \times d_{model}}$ where $N_u$ is the number of sparse features of {\itshape User Profile}. {\itshape Item Profile} can be represented by $\textbf{X}^I \in \mathbb{R}^{N_i \times d_{model}}$ where $N_i$ is the number of sparse features of {\itshape Item Profile}. {\itshape User Behavior} can be represented by $\textbf{S} = [\textbf{b}_1;...;\textbf{b}_i;...;\textbf{b}_N] \in \mathbb{R}^{N \times d_{model}}$ where $N$ is the number of users' historical behaviors and $\textbf{b}_i$ is the embedding of the $i$-th behavior.

\subsubsection{Multiple Layer Perceptron (MLP)}
First, embeddings of sparse features from {\itshape User Profile}, {\itshape Item Profile} and {\itshape User Behavior} are concatenated, flattened and then fed into MLP with the activation function such as RELU. The softmax function is used at last to predict the probability of the user clicking on the target item.

\subsubsection{Loss Function}
The negative log-likelihood function is widely used in CTR models, which is usually defined as:
\begin{equation} \label{eq:loss function}
\begin{split}
L = - \frac{1}{N} \sum_{(x, y) \in \mathbb{D}} (y\,\log p(x) + (1 - y)\,\log(1 - p(x)))
\end{split}
\end{equation}
where $\mathbb{D}$ is the training dataset, $x$ is represented by $[\textbf{X}^U,\textbf{X}^I,\textbf{S}]$ is the input of the network, $y \in \{0, 1\}$ represents whether the user clicked the item and $p(\cdot)$ is the final output of the network which represents the prediction probability that the user clicks the item.

\begin{figure*}[ht]
\begin{flushleft}
\includegraphics[scale=0.366]{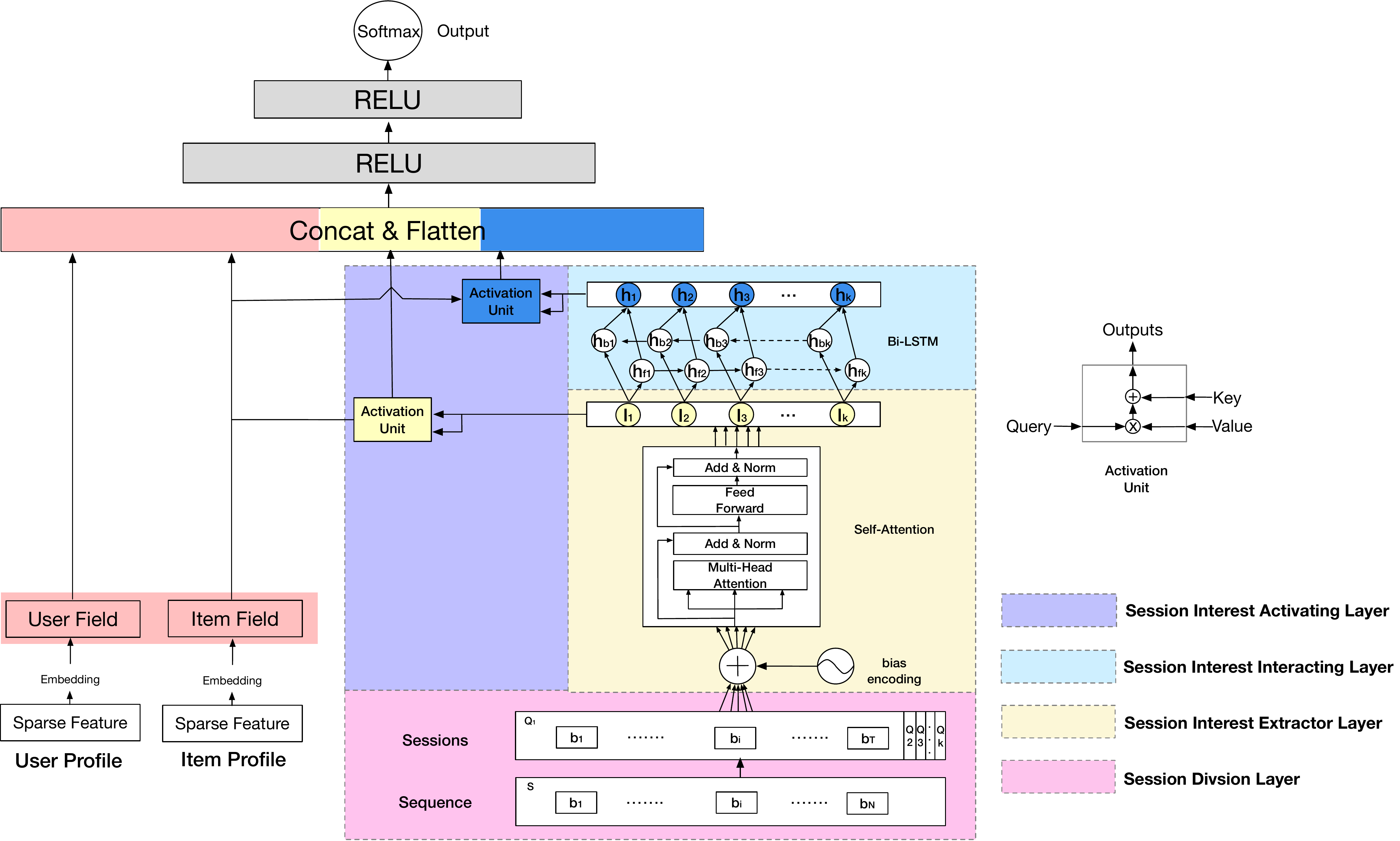}
\end{flushleft}
\caption{The overview of our proposed model DSIN. Overall, before MLP layers, DSIN has two main components. One is sparse features and the other processes the user behavior sequence. From the bottom up, the user behavior sequence $\textbf{S}$ is first divided into sessions $\textbf{Q}$, which are then added with bias encoding and extracted into session interests $\textbf{I}$ with self-attention. With Bi-LSTM, we mix session interests $\textbf{I}$ with contextual information as hidden states $\textbf{H}$. Both Vectors of session interests $\textbf{I}$ and hidden states $\textbf{H}$ activated by the target item and embedding vectors of {\itshape User Profile} and {\itshape Item Profile} are concatenated, flattened and then fed into MLP layers for the final prediction.
}
\label{fig:DSIN}
\end{figure*}

\subsection{Model Overview}
In recommnder systems, users' behavior sequences consist of multiple historical sessions. Users show varying interests in different sessions. Also, users' session interests are sequentially related to each other. DSIN is proposed to extract users' session interest in each session and capture the sequential relationship of session interests. 

As shown in figure \ref{fig:DSIN}, DSIN consists of two parts before MLP. One is the embedding vectors transformed from {\itshape User Profile} and {\itshape Item Profile}. The other models {\itshape User Behavior} and has four layers from the bottom up: (1) session division layer partitions users' behavior sequence into sessions; (2) session interest extractor layer extracts users' session interests; (3) session interest interacting layer captures the sequential relationship among session interests; (4) session interest activating layer applies the local activation unit to users' session interests w.r.t the target item. Finally outputs of session interest activating layer and embedding vectors of {\itshape User Profile} and {\itshape Item Profile} are fed into MLP for the final prediction. In the following sections we introduce these four layers in the latter part in detail.

\subsubsection{Session Division Layer}
To extract more precise users' session interests, we divide users' behavior sequences $\textbf{S}$ into sessions $\textbf{Q}$, where the $k$-th session $\textbf{Q}_k = [\textbf{b}_1;...;\textbf{b}_i;...;\textbf{b}_T] \in \mathbb{R}^{T \times d_{model}}$, $T$ is the number of behaviors we keep in the session and $\textbf{b}_i$ is users' $i$-th behavior in the session. The segmentation of users' sessions exists between adjacent behaviors whose time interval is more than 30 minutes followed by \cite{Mihajlo:airbnb}. 

\subsubsection{Session Interest Extractor Layer}
Behaviors in the same session are strongly related to each other. Besides, users' casual behaviors in the session deviate the session interest from its original expression. To capture the  inner relationship between behaviors in the same session and decrease the effect of those unrelated behaviors, we employ multi-head self-attention \cite{Vaswani:Transformer} mechanism in each session. We also make some improvements in the self-attention mechanism to achieve our goal better.

\paragraph{Bias Encoding.} To make use of the order relations of the sequence, self-attention mechanism applies positional encoding to the input embeddings. Furthermore, the order relations of sessions and the bias existed in different representation subspaces also need to be captured. Thus, we propose bias encoding $\textbf{BE} \in \mathbb{R}^{K \times T \times d_{model}}$ on the basis of positional encoding, where each element in $\textbf{BE}$ is defined as follows:
\begin{equation} 
\begin{split}
\textbf{BE}_{(k, t, c)} &= \textbf{w}_{k}^{K} + \textbf{w}_{t}^{T} + \textbf{w}_{c}^{C}
\end{split}
\end{equation}
where $\textbf{w}^K \in \mathbb{R}^{K}$ is the bias vector of the session, $k$ is the index of sessions, $\textbf{w}^T \in \mathbb{R}^{T}$ is the bias vector of the position in the session, $t$ is the index of the behavior in sessions, $\textbf{w}^C \in \mathbb{R}^{d_{model}}$ is the bias vector of the unit position in the behavior embedding and $c$ is the index of the unit in the behavior embedding. After added with bias encoding, users' behavior sessions $\textbf{Q}$ are updated as follows:
\begin{equation} 
\begin{split}
\textbf{Q} = \textbf{Q} + \textbf{BE}\\
\end{split}
\end{equation}

\paragraph{Multi-head Self-attention.} In recommender systems, users' click behaviors are influenced by various factors (e.g. colors, styles and price) \cite{zhou2018atrank}. Mulit-head self-attention can capture relationship in different representation subspaces. Mathematically, let $\textbf{Q}_k = [\textbf{Q}_{k1};...;\textbf{Q}_{kh};...;\textbf{Q}_{kH}]$ where $\textbf{Q}_{kh} \in \mathbb{R}^{T \times d_{h}}$ is the $h$-th head of $\textbf{Q}_k$, $H$ is the number of heads and $d_{h} = \frac{1}{h}d_{model}$. The output of $\textbf{head}_{h}$ is calculated as follows:
\begin{equation}
\begin{split}
\textbf{head}_h &= \mbox{Attention}(\textbf{Q}_{kh}\textbf{W}^Q, \textbf{Q}_{kh}\textbf{W}^K, \textbf{Q}_{kh}\textbf{W}^V)\\
&= \mbox{softmax}(\frac{\textbf{Q}_{kh}\textbf{W}^Q{\textbf{W}^{K}}^T\textbf{Q}_{kh}^{T}}{\sqrt{d_{model}}})\textbf{Q}_{kh}\textbf{W}^V\\
\end{split}
\end{equation}
where $\textbf{W}^Q, \textbf{W}^K, \textbf{W}^Q$ are linear matrices. Then vectors of different heads are concatenated and then fed into a feed-forward network:
\begin{equation} 
\begin{split}
\textbf{I}^Q_k = \mbox{FFN}(\mbox{Concat}(\textbf{head}_{1}, ..., \textbf{head}_{H})\textbf{W}^O)\\
\end{split}
\end{equation}
where $\mbox{FFN}(\cdot)$ is the feed-forward network and $\textbf{W}^O$ is the linear matrix. We also conduct residual connections and layer normalization successively. Users' $k$-th session interest $\textbf{I}_k$ is calculated as follows:
\begin{equation}
\begin{split}
\textbf{I}_k = \mbox{Avg}(\textbf{I}^Q_k)
\end{split}
\end{equation}
where $\mbox{Avg}(\cdot)$ is the average pooling. Note that weights are shared in the self-attention mechanism of different sessions.

\subsubsection{Session Interest Interacting Layer}
Users' session interests hold sequential relations with contextual ones. Modeling the dynamic changes enriches the representation of the session interests. Bi-LSTM \cite{Graves:bilstm} is excellent at capturing sequential relations and naturally applied on modeling the interaction of session interests in DSIN. LSTM \cite{Hochreiter:LSTM} memory cell is implemented as follows:
\begin{equation} \label{eq:bi-lstm}
\begin{aligned}
\textbf{i}_t &= \sigma(\textbf{W}_{xi}\textbf{I}_t + \textbf{W}_{hi}\textbf{h}_{t-1} + \textbf{W}_{ci}\textbf{c}_{t-1} + \textbf{b}_i)\\
\textbf{f}_t &= \sigma(\textbf{W}_{xf}\textbf{I}_t + \textbf{W}_{hf}\textbf{h}_{t-1} + \textbf{W}_{cf}\textbf{c}_{t-1} + \textbf{b}_f)\\
\textbf{c}_t &= \textbf{f}_{t}\textbf{c}_{t-1} + \textbf{i}_{t}\tanh(\textbf{W}_{xc}\textbf{I}_t + \textbf{W}_{hc}\textbf{h}_{t-1} + \textbf{b}_{c})\\
\textbf{o}_t &= \sigma(\textbf{W}_{xo}\textbf{I}_t + \textbf{W}_{ho}\textbf{h}_{t-1} + \textbf{W}_{co}\textbf{c}_{t} + \textbf{b}_o)\\
\textbf{h}_t &= \textbf{o}_{t}\tanh(\textbf{c}_t)\\
\end{aligned}
\end{equation}
where $\sigma(\cdot)$ is the logistic function, and $\textbf{i}$, $\textbf{f}$, $\textbf{o}$ and $\textbf{c}$ are the input gate, forget gate, output gate and cell vectors which have the same size as $\textbf{I}_t$. Shapes of weight matrices are indicated with the subscripts. Bi-direction means that there exist forward and backward RNNs, and the hidden states $\textbf{H}$ are calculated as follows:
\begin{equation}
\begin{split}
\textbf{H}_t = \overrightarrow{\textbf{h}_{ft}} \oplus \overleftarrow{\textbf{h}_{bt}}
\end{split}
\end{equation}
where $\overrightarrow{\textbf{h}_{ft}}$ is the hidden state of the forward LSTM and $\overleftarrow{\textbf{h}_{bt}}$ is the hidden state of the backward LSTM.

\subsubsection{Session Interest Activating Layer}
Users' session interests more related to the target item have greater impacts on whether the user will click the target item. The weights of users' session interests need to be reallocated w.r.t. the target item. Attention mechanism \cite{Dzmitry:Attention} conducts soft alignment between source and the target and has been proved effective as a weight allocation mechanism. The adaptive representation of session interests w.r.t. the target item is calculated as follows:
\begin{equation}
\begin{split}
a^I_k &= \frac{\exp(\textbf{I}_k\textbf{W}^I\textbf{X}^I))}{\sum_{k}^{K} \exp(\textbf{I}_k\textbf{W}^I\textbf{X}^I)} \\
\textbf{U}^I &= \sum_{k}^{K} a^I_k\textbf{I}_k\\
\end{split}
\end{equation}
where $\textbf{W}^I$ has the corresponding shape. Similarly, the adaptive representation of session interests mixed with contextual information w.r.t. the target item is calculated as follows:
\begin{equation}
\begin{split}
a^H_k &= \frac{\exp(\textbf{H}_k\textbf{W}^H\textbf{X}^I))}{\sum_{k}^{K} \exp(\textbf{H}_k\textbf{W}^H\textbf{X}^I)} \\
\textbf{U}^H &= \sum_{k}^{K} a^H_k\textbf{H}_k
\end{split}
\end{equation}
where $\textbf{W}^H$ has the corresponding shape. Embedding vectors of $User$ $Profile$ and $Item$ $Profile$, $\textbf{U}^I$ and $\textbf{U}^H$ are concatenated, flattened and then fed into the MLP layer.

\section{Experiments}
In this section, we first introduce experiment datasets, competitors and evaluation metric. Then we compare our proposed DSIN with competitors and analyse the results. We further discuss the effectiveness of critical technical designs in DSIN empirically. 

\subsection{Datasets}
\subsubsection{Advertising Dataset}
Advertising Dataset\footnote{https://tianchi.aliyun.com/dataset/dataDetail?dataId=56} is a public dataset released by Alimama, an online advertising platform in China. It contains 26 million records from ad display/click logs of 1 million users and 800 thousand ads in 8 days. Logs from $2017$-$05$-$06$ to $2017$-$05$-$12$ are for training and logs from $2017$-$05$-$13$ are for testing. Users' recent 200 behaviors are also recorded in logs.

\subsubsection{Recommender Dataset}
To verify the effectiveness of DSIN in the real-world industrial applications, we conduct experiments on the recommender dataset of Alibaba. This dataset contains 6 billion display/click logs of 100 million users and 70 million items in 8 days. Logs from $2018$-$12$-$13$ to $2018$-$12$-$19$ are for training and logs from $2018$-$12$-$20$ are for testing. Users' recent 200 behaviors are also recorded in logs. 

\subsection{Competitors}
\begin{itemize}
\item \textbf{YoutubetNet}. YoutubeNet \cite{Paul:YoutubeNet} is a technically designed model which uses users' watching video sequence for video recommendation in Youtube. It treats users' historical behaviors equally and utilizes average pooling operation. We also experiment with YoutubeNet without $User$ $Behavior$ to verify the effectiveness of historical behaviors.
\item \textbf{Wide\&Deep}. Wide\&Deep \cite{Cheng:WDL} is a CTR model with both memorization and generalization. It contains two parts: wide model of memory and deep model of generalization.
\item \textbf{DIN}. Deep Interest Network \cite{Zhou:DIN} fully exploits the relationship between users' historical behaviors and the target item. It uses attention mechanism to learn the representation of users’ historical behaviors w.r.t. the target item.
\item \textbf{DIN-RNN}. DIN-RNN has a similar structure as DIN, except that we use the hidden states of Bi-LSTM, which models users' historical behaviors and learns the contextual relationship.
\item \textbf{DIEN}. DIEN \cite{zhou2018deep} extracts latent temporal interests from user behaviors and models interests evolving process. Auxiliary loss makes hidden states more expressive to represent latent interests and AUGRU models the specific interest evolving processes for different target items.
\end{itemize}
\subsection{Metrics}
AUC (Area Under ROC Curve) reflects the ranking ability of the model. It is defined as follows:
\begin{equation}
\begin{split}
\mbox{AUC} = \frac{1}{m^{+}m^{-}}\sum_{x^{+}\in D^{+}}\sum_{x^{-}\in D^{-}}(I(f(x^{+})>f(x^{-}))))
\end{split}
\end{equation}
where $D^{+}$ is the collection of all positive examples, $D^{-}$ is the collection of all negative examples, $f(\cdot)$ is the result of the model's prediction of the sample x and $I(\cdot)$ is the indicator function.\\
\subsection{Results on the advertising and recommender Datasets}
\begin{figure*}[htbp]
\begin{center}
\subfigure{
\includegraphics[scale=0.08]{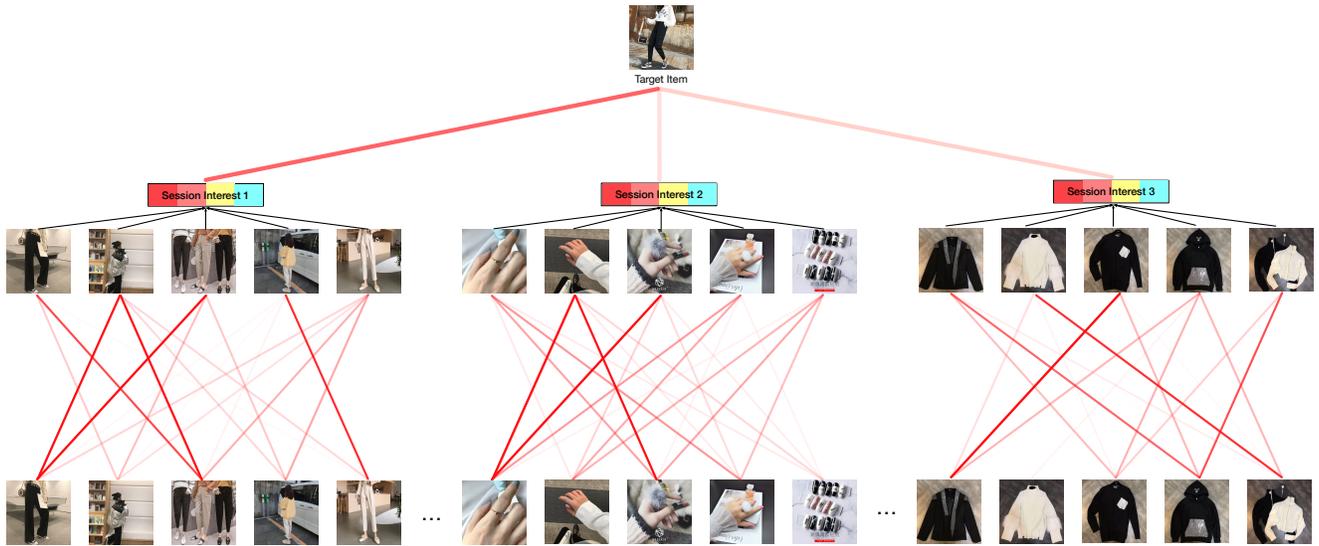}
}
\end{center}
   \caption{This figure visualizes the attention weights in the self-attention mechanism at the lower half and the activation unit work at the upper half in DSIN. Note that the attention weights in the self-attention mechanism are the sum of that in each head. Also, the darker the line color, the greater the weight. 
}
\label{figure3}
\end{figure*}

\begin{table}
\centering
\begin{threeparttable}
\begin{tabular}{lrr}  
\toprule
Model & Advertising & Recommender\\
\midrule
YoutubeNet-NO-UB$^a$ & 0.6239 & 0.6419  \\
YoutubeNet & 0.6313 & 0.6425  \\
DIN-RNN & 0.6319 & 0.6435 \\
Wide\&Deep & 0.6326 & 0.6432 \\
DIN & 0.6330 & 0.6459 \\
DIEN & 0.6343 & 0.6473 \\
\hline
DSIN-PE$^b$ & 0.6357 & 0.6494 \\
DSIN-BE-NO-SIIL$^c$ & 0.6365 & 0.6499 \\
\textbf{DSIN-BE}$^d$ & \textbf{0.6375} & \textbf{0.6515} \\
\bottomrule
\end{tabular}
\begin{tablenotes}
\item[a] YoutubeNet without {\itshape User Behavior}.
\item[b] DSIN with positional encoding.
\item[c] DSIN with bias encoding and without session interest interacting layer and the corresponding activation unit.
\item[d] DSIN with bias encoding.
\end{tablenotes}
\end{threeparttable}
\caption{Results (AUC) on the advertising and recommender dataset}
\label{tab:booktabs}
\end{table}

Results on the advertising dataset and recommender dataset are shown in Table 1. YoutubeNet performs better than YoutubeNet-No-User-Behavior owing to {\itshape User Behavior}, while Wide\&Deep gets the betters result due to combining the memorization of wide part. DIN improves AUC obviously by activating {\itshape User Behavior} w.r.t. the target item. Especially, the results of DIN-RNN in both datasets are worse than those of DIN due to the discontinuity of users' behavior sequences. DIEN obtains better results while auxiliary loss and specially designed AUGRU lead to deviating from the original expression of behaviors. DSIN gets the best results on both datasets. It extracts users' historical behaviors into session interests and models the dynamic evolving procedure of session interests, both of which enrich the representation of the user. The local activation unit helps obtain the adaptive representation of users' session interests w.r.t. the target item.


\subsection{Further Discussion}
\subsubsection{Effect of Multiple Sessions}
As shown in Table \ref{tab:booktabs}, results show that DIN-RNN performs worse than DIN while DSIN-BE performs better than DSIN-BE-NO-SIIL. The difference between each pair is only the sequence modeling. \cite{Zhou:DIN} explains that rapid jumping and sudden ending over behaviors causes the sequence data of user behaviors to seem to be noisy. It will lead to information loss in the process of information transmission in RNNs and further confuse the representation of users' behavior sequences. 
While in DSIN, we partition users' behavior sequences into multiple sessions for the following two reasons: (i) users' behaviors are generally homogeneous in each session; (ii) users' session interests follow a sequential pattern and are more suitable for sequence modeling. Both improve the performance of DSIN.

\subsubsection{Effect of Session Interest Interacting Layer}
As shown in Table 1, we conduct comparative experiments with DSIN-BE and DSIN-BE-NO-SIIL, where DSIN-BE performs better. With session interest interacting layer, users' session interests are mixed with contextual information and become more expressive, which improve the performance of DSIN.

\subsubsection{Effect of Bias Encoding}
As shown in Table 1, we conduct comparative experiments with DSIN-BE and DSIN-PE, where DSIN-BE performs better. Different from the two-dimensional positional encoding, the bias of users' sessions is also captured. Empirically, bias encoding successfully captures the order information of sessions and improves the performance of DSIN.

\subsubsection{Visualization of Self-attention and the Activation Unit}
As shown in figure \ref{figure3}, we visualize the attention weights in the the local activation unit and self-attention mechanism.
To illustrate the effect of self-attention, we take the first session for example. The user mainly browses trouser-related items and occasionally coat-related items. We can observe that weights of trouser-related items are generally high. After self-attention, most representations of trouser-related behaviors are reserved and extracted into the user's interest in this session.
Besides, the local activation unit works by making users' session interests related to the target item more prominent. In this case, the target item is a black trouser. The user's trouser-related session interest is assigned greater weight and has more influence on the final prediction. While the session 3 is coat-related, the user's color preference to black in it is also helpful to rank the trouser.

\section{Conclusion}
In this paper, we provide a novel perspective on the CTR prediction task, where users' sequential behaviors consist of multiple historical sessions. User behaviors are highly homogeneous in each session and heterogeneous in different sessions.
Base on these observations, Deep Session Interest Network (DSIN) is proposed. We first use the self-attention mechanism with bias encoding to extract the user's interest of each session. Then we apply Bi-LSTM to capture the sequential relation of contextual session interests. We employ the local activation unit to aggregate the user's different session interest representations with regard to the target item at last. Experiment results demonstrate the effectiveness of DSIN both on advertising and recommender datasets.
In the future, we will pay attention to utilizing knowledge graph as prior knowledge to explain users' historical behaviors for better CTR prediction.
\newpage
\bibliographystyle{named}
\bibliography{ijcai19}

\end{document}